\documentclass[aps,amsmath,amssymb,superscriptaddress,preprint,nofootinbib]{revtex4-1}

\usepackage{graphicx}
\usepackage{dcolumn}
\usepackage{bm}

\usepackage{hyperref}

\usepackage{color}
\usepackage{amsfonts}
\usepackage{amssymb}
\usepackage{amsmath}

\begin{document}

\title{S-matrix formulation of thermodynamics with N-body scatterings}
\author{Pok Man Lo}
\affiliation{Institute of Theoretical Physics, University of Wroclaw,
PL-50204 Wroc\l aw, Poland}
\affiliation{Extreme Matter Institute EMMI, GSI, D-64291 Darmstadt, Germany}
\email{pmlo@gsi.de}

\begin{abstract}
We apply a phase space expansion scheme to incorporate the N-body scattering processes in the S-matrix formulation of statistical mechanics.
A generalized phase shift function suitable for studying the thermal contribution of $N \rightarrow N$ processes is motivated and examined in various models.
Using the expansion scheme, we revisit how the hadron resonance gas model emerges from the S-matrix framework, and
consider an example of structureless scattering in which the phase shift function can be exactly worked out. 
Finally we analyze the influence of dynamics on the phase shift function in a simple example of 3- and 4-body scattering.
\end{abstract}

\maketitle

\section{Introduction}

The S-matrix formulation of statistical mechanics by Dashen, Ma and Bernstein~\cite{dmb} allows the computation of the grand canonical potential in terms of scattering matrix elements. 
The approach has been applied to study the thermal properties of an interacting gas of hadrons~\cite{Venugopalan:1992hy}.
Relying on the empirical data of the scattering phase shifts, the contributions from both the low-lying resonances and the purely repulsive channels are consistently included in the description of thermodynamics. In particular, resonances with a large width like the $\sigma$- and $\kappa$-mesons can be appropriately treated within this method~\cite{sigma, kappa}.

The use of the empirical phase shifts makes the approach to some extent model independent. However, this necessarily restricts the application of the approach to the 2-body sector and to the channels and energy range in which experimental data are available. The latter may be solved by complementary model calculations of the relevant S-matrix elements. For the limitation to 2-body scatterings, while it may be justified in the low density case, where ($N > 2$)-body scatterings are expected to be rare, inclusion of higher N-body scatterings is essential for checking the validity of the expansion and for extending the approach to describe a dense medium.

The need for the systematic inclusion of the N-body interactions in describing the thermodynamics is common for many theoretical approaches. In the functional approach, 
sophisticated truncation schemes~\cite{Maris:2003vk} have been devised to include a subset of N-point functions for the calculation of the 2-point function. 
Treatment of the ($N > 2$)-point functions themselves as the object of interest, though not at the same level of sophistication, has begun~\cite{Popovici:2011yz}.

On the other hand, the S-matrix formulation represents an alternative expansion scheme of the thermodynamic potential, involving only the on-shell S- or T-matrix elements~\cite{dmb}.
Unfortunately not much is known about the higher N-body contribution. It is therefore a useful exercise to explore the qualitative behaviors of these correction terms, even in the context of some simplified models.

The S-matrix framework discussed here is flexible enough to receive inputs from field theoretical models, potential models or experiments.
Hence it is more practical to separate the following two issues. First, the problem of searching for a theoretical model to describe the physical S-matrix. Second, the exploration of the influences on macroscopic quantities based on these S-matrix elements. In this work, we shall discuss the latter issue and make some efforts to elucidate the expansion of the N-body trace in terms of the Lorentz invariant phase space. This offers important insights into how the N-body scattering processes enter the thermodynamics. Building on this, 
we work towards applying the S-matrix approach beyond the 2-body setting within some simplified model amplitudes.

This paper is organized as follows. In the next section, we review some of the well-known results of the S-matrix formalism in the context of quantum mechanics. A phase space expansion scheme for handling the 2-body trace is introduced and the definition of a generalized phase shift function, suitable for describing an $N \rightarrow N$ scattering process, is motivated.
In Sec. 3, we consider some applications of the scheme. This includes the demonstration of how the hadron resonance gas (HRG) model emerges naturally from the framework. Also a structureless scattering model will be considered which illustrates how the N-body phase shift can be exactly worked out. 
In sec. 4 we explore the influence of dynamics on the phase shift function via the quantum amplitudes in a simple example of 3- and 4-body scattering.
In sec. 5 we present the conclusion.

\section{S-matrix approach to thermodynamics}
\label{sec:2}

Our starting point is the S-matrix formulation of statistical mechanics by Dashen, Ma and Bernstein~\cite{dmb}. It states that the interacting part of the grand canonical potential can be expressed in terms of the S-matrix, 
which describes the scattering processes within a thermodynamical system~\footnote{For simplicity, we do not tackle the problem of quantum statistics in this work. To do this, one needs to perform proper symmetrization or anti-symmetrization of the states in the trace.}:

\begin{align}
  \label{eq:main}
    \Delta \ln Z &= \int d E \,  e^{- \beta E} \, \frac{1}{4 \pi i} \, {\rm tr} \, \left[ \, S^{-1} \, \overleftrightarrow{\frac{\partial}{\partial E}} \, S \, \right]_c.
\end{align}

\noindent The subscript c here corresponds to taking only the connected contributions in the trace. Furthermore, as will become evident in the discussion, it is useful to rewrite the derivative operator in the following form:

\begin{align}
  \begin{split}
    {\rm tr} \, \left[ S^{-1} \, \overleftrightarrow{\frac{\partial}{\partial E}} \, S \right]_c &= {\rm tr} \, \left[ \, S^{-1} \, ({\frac{\partial}{\partial E}} \, S) - ({\frac{\partial}{\partial E}} \, S^{-1} ) \, S \, \right]_c \\
    &=  2 \, \frac{\partial}{\partial E} \, {\rm tr} \, \left[ \,  \ln S \, \right]_c.
  \end{split}
\end{align}

\subsection{Beth-Uhlenbeck result revisited}

We first discuss the S-matrix approach in the context of quantum mechanics. If we simply replace the S-matrix operator by

\begin{align}
  S \rightarrow e^{\, 2 i \delta_E},
\end{align}

\noindent we obtain the well-known result of Beth and Uhlenbeck~\cite{Beth:1937zz}.

\begin{align}
     \label{eq:BU}
     \Delta \ln Z  &= \int d E \, e^{- \beta E} \times \frac{1}{\pi } \, \frac{\partial}{\partial E} \, {\rm tr} \, \left( \delta_E  \right).
\end{align}

A more formal way to derive this result is to express the S-matrix operator in terms of the scattering Green's function~\cite{How:2010zz}. Consider the decomposition of the Hamiltonian into

\begin{align}
    H &= H_0 + V,
\end{align}

\noindent the non-interacting and the full Green's functions are given by

\begin{align}
  \begin{split}
    G_0 &= \frac{1}{E-H_0 + i \epsilon} \\
    G &= \frac{1}{E-H + i \epsilon}.
  \end{split}
\end{align}

\noindent We shall show that the S-matrix operator appeared in Eq.~\eqref{eq:main} can be expressed by

\begin{align}
  S = G^*_0 \, {G^*}^{-1} \, G \, {G_0}^{-1}.
\end{align}

\noindent This expression follows from the Lippmann-Schwinger equation. To see that, we recast the formula into a different form:

\begin{align}
  \begin{split}
    S &= G^*_0 \, {G^*}^{-1} \, G \, {G_0}^{-1} \\
    &= G_0^* \, \left( I - 2 i \epsilon \times G \right) \, {G_0}^{-1} \\
    &= G_0^* {G_0}^{-1} - 2 i \epsilon \times G^*_0 \times  \left( G-G_0 \right) \, {G_0}^{-1} \\
    &= I - 2 i \epsilon \times G^*_0 G_0 \times V \times G {G_0}^{-1} \\
    &= I + \left( G_0-G^*_0 \right) \times V G G_0^{-1}.
  \label{eq:LS-1}
  \end{split}
\end{align}

\noindent Recall the expression of T-matrix from the Lippmann-Schwinger equation

\begin{align}
  \begin{split}
    T &= V + V G V \\
    G &= G_0 + G_0 V G,
  \end{split}
\end{align}

\noindent which gives

\begin{align}
  T &= V G G_0^{-1}.
  \label{eq:LS-2}
\end{align}

\noindent Combining equations~\eqref{eq:LS-1} and~\eqref{eq:LS-2}, we obtain

\begin{align}
  \begin{split}
    S &= G^*_0 \, {G^*}^{-1} \, G \, {G_0}^{-1} \\
    &= I - 2 \pi i \times \delta(E-H_0) \times T, 
  \end{split}
\end{align}

\noindent which matches the standard definition of the S-matrix for the scattering theory. In performing the trace the operator will be surrounded by quantum states, the proportionality to $\delta(E-H_0)$ means that only the on-shell matrix elements are involved, and hence the replacement of $ S \rightarrow e^{\, 2 i \delta_E} $ is valid. It remains to show its connection to the thermodynamic potential.

Consider the free partition function

\begin{align}
  \begin{split}
    Z_0 &= {\rm tr} \, e^{-\beta H_0} \\
     &= \int d E \, e^{- \beta E} \frac{1}{2 \pi } \, {\rm tr} \, \left[ 2 \, \pi \times \delta(E-H_0)  \right]\\
     &= \int d E \, e^{- \beta E} \frac{1}{2 \pi i}  (-1) \times  {\rm tr} \, \left[ G_0 - G_0^* \right].
  \end{split}
\end{align}

\noindent The corresponding result for the interacting part of the logarithm of the full partition function reads

\begin{align}
  \begin{split}
    \Delta \ln Z &= \int d E \, e^{- \beta E} \frac{1}{2 \pi i} \times \\
    &(-1) \times  {\rm tr} \, \left[ (G-G^*) - (G_0 - G_0^*) \right]_c.
  \end{split}
\end{align}

\noindent It is then straightforward to verify that 

\begin{align}
  \begin{split}
    &\frac{\partial}{\partial E} \, {\rm tr} \, \left[ \, \ln S \, \right ]_c \\
    &=  \frac{\partial}{\partial E} \, {\rm tr} \, \left[ \, \ln \, G^*_0 \, {G^*}^{-1} \, G \, {G_0}^{-1} \, \right]_c \\
    &= (-1) \times  {\rm tr} \, \left[ (G-G^*) - (G_0 - G_0^*) \right]_c,
  \end{split}
\end{align}

\noindent and finally reaching the result stated in Eq.~\eqref{eq:BU}. This concludes our alternative derivation of the Beth-Uhlenbeck result via the scattering Green's functions.

\subsection{Generalized phase shift function $Q(M)$}

If the empirical phase shift for a 2-body interaction is measured in an experiment, the data can be used directly in Eq.~\eqref{eq:BU} to obtain its contribution to the thermodynamics~\cite{Venugopalan:1992hy, Weinhold:1997ig, sigma, kappa}.
When no such data is available, it is necessary to obtain the relevant S-matrix element from a model.
For a more general application of the approach, we consider the case where the S-matrix elements are obtained within a field theoretical model.

Staying within the 2-body sector, we start from the elementary definition of the T-matrix

\begin{align}
  S_{\rm QFT} = I + i T_{\rm QFT},
\end{align}

\noindent and relate the T-matrix element to a quantum field amplitude via~\cite{How:2010zz}

\begin{align}
  \langle k_1^\prime k_2^\prime \vert \, i T_{\rm QFT} \, \vert k_1 k_2 \rangle =& \, (2 \pi)^4 \times \delta_E^4  \times i \mathcal{M}_{k_1^\prime, k_2^\prime; k_1 k_2},
\end{align}

\noindent where

\begin{align}
  \begin{split}
    \delta_E^4 \equiv& \, \delta(E-E_1-E_2) \, \times  \delta^{3}(\vec{k^\prime_1} + \vec{k^\prime_2}-\vec{k_1}-\vec{k_2}).
  \end{split}
\end{align}

\noindent The amplitude $i \mathcal{M}_{k_1^\prime, k_2^\prime; k_1 k_2}$ can be constructed using the standard Feynman rules~\footnote{The amplitude usually involves a ladder sum of a set of tree-level diagrams.}.

To facilitate the evaluation of the 2-body trace in Eq.~\eqref{eq:main}, we introduce the following shorthand notations

\begin{align}
  \begin{split}
    \int (dk) \, (\, \cdots) \rightarrow& \int \frac{d^3 p_1}{(2 \pi)^3} \frac{1}{2 E_1}\frac{d^3 p_2}{(2 \pi)^3} \frac{1}{2 E_2} \, (\, \cdots) \\
    \int d \phi_2 \, (\,\cdots) \rightarrow& \int \frac{d^3 p_1}{(2 \pi)^3} \frac{1}{2 E_1}\frac{d^3 p_2}{(2 \pi)^3} \frac{1}{2 E_2} \times \\
    &(2 \pi)^4 \, \delta^4(P_I-\sum_i p_i)
    \, (\,\cdots). 
  \end{split}
\end{align}

\noindent Following the discussion in Ref.~\cite{How:2010zz}, we consider a phase space expansion for evaluating the 2-body trace:

\begin{align}
\begin{split}
    &{\rm tr}_2 \, \ln S_{\rm QFT}  \\
    &= \int (d \, k) \,  \langle k_1 k_2 \vert \, \ln S_{\rm QFT} \, \vert k_1 k_2 \rangle \\
    &= \int (d \, k) \, \sum_l (-1) \frac{(-i)^l}{l}   \langle k_1 k_2 \vert \, T_{\rm QFT}^l \, \vert k_1 k_2 \rangle \\
    &\approx \sum_l (-1) \frac{(-i)^l}{l}  \int (d \, k) (d \, k^{(1)}) (d \, k^{(2)}) \cdots  (d \, k^{(l-1)}) \, \times  \\
    &\hspace{0.25cm} \langle k_1 k_2 \vert \, T_{\rm QFT} \, \vert k_1^{(1)} k_2^{(1)} \rangle \, \times \langle k_1^{(1)} k_2^{(1)} \vert \, T_{\rm QFT} \, \vert k_1^{(2)} k_2^{(2)} \rangle \, \times  \\
    &\cdots \langle k_1^{(l-1)} k_2^{(l-1)} \vert \, T_{\rm QFT} \, \vert k_1 k_2 \rangle  \\
    &= \sum_l (-1) \frac{(-i)^l}{l}  \int d \phi_2 \, d \phi_2^{(1)} \, d \phi_2^{(2)} \cdots d \phi_2^{(l-1)} \, \times \\
    &\hspace{0.25cm} \mathcal{M}_{k_1, k_2; k_1^{(1)} k_2^{(1)}} \, \mathcal{M}_{k_1^{(1)} k_2^{(1)}; k_1^{(2)} k_2^{(2)}} \, \cdots \, \times  \\
    &\mathcal{M}_{k_1^{(l-1)} k_2^{(l-1)}; k_1 k_2} \, \times \left[V \frac{d^3 P}{(2 \pi)^3}\right] \\
    &\approx \sum_l (-1) \frac{(-i)^l}{l}  \langle \int d \phi_2 \, \mathcal{M} \rangle^l  \, \left[V \frac{d^3 P}{(2 \pi)^3}\right]  \\
    &= \ln \left( 1 + i \, \langle \int d \phi_2 \, \mathcal{M} \rangle \right) \, \times \left[V \frac{d^3 P}{(2 \pi)^3}\right].
\end{split}
\tag{$\star$}
\end{align}

\noindent Here we highlight some key steps in the derivation. The volume factor $V$ comes from the redundant 3-dimensional Dirac-delta function in closing the chain of the resolution of the identity from $(k_1^{(l-1)} k_2^{(l-1)}) \rightarrow (k_1 k_2)$, i.e., 

\begin{align}
  V = (2 \pi)^3 \times \delta^{3}(\vec{k_1} + \vec{k_2}-\vec{k_1}-\vec{k_2}).
\end{align}

\noindent Inserting 

\begin{align}
\int \frac{d^3 P}{(2 \pi)^3} \, (2 \pi)^3 \times \delta^{3}(\vec{P} -\vec{k_1}-\vec{k_2})
\end{align}

\noindent allows to complete the leftover integral from

\begin{align}
  \int (dk) \, 2 \pi \times \delta(E-E_1-E_2) (\, \cdots)
\end{align}

\noindent to 

\begin{align}
\int d\phi_2 \, (\, \cdots).
\end{align}

Furthermore, the first approximation sign corresponds to the restriction to $2 \rightarrow 2$ processes with no change in particle identities, i.e., elastic scattering. The second one corresponds to a factorization approximation. In case of a constant amplitude, or an amplitude that depends only on the invariant mass $ M = \sqrt{E^2-\vec{P}^2} $, this approximation is exact. Generally the various phase space integrals are coupled and cannot be factorized. The notation $\langle \cdots \rangle$ serves as a reminder of this fact.

With such an expression, we reach the following formula for the thermodynamic pressure due to the 2-body interaction with quantum amplitude $i \mathcal{M}$:

\begin{align}
  \begin{split}
    \label{eq:Q_def}
    (\Delta \ln Z) &= V \, \int \frac{d^3 P}{(2 \pi)^3} \,  \frac{ d M}{2 \pi} \, e^{- \beta \sqrt{P^2 + M^2}} \, B(M)  \\
    B(M) &\equiv 2 \, {\frac{\partial}{\partial M}} Q(M) \\
    Q(M) &\equiv \frac{1}{2} \, {\rm Im} \, \ln \left[ 1 + i \, \langle \int d \phi_2 \, \mathcal{M} \rangle \right].
  \end{split}
\end{align}

\noindent We take this chance to introduce a phase shift function $Q(M)$ and a corresponding effective spectral function $B(M)$. As we shall see in the coming sections, ${Q}$ is a suitable generalization of the phase shift in 2-body case for discussing $N \rightarrow N$ processes, after replacing the integral over the 2-body phase space $\phi_2 $ with an N-body one.

\subsection{Expansion in terms of T-matrix}

It may be helpful to express the previous results in terms of the T-matrix:

\begin{align}
  \begin{split}
    &{\rm tr} \, \left[ S^{-1} \, \overleftrightarrow{\frac{\partial}{\partial E}} \, S \right]_c \\
    &= \, {\rm tr} \, \left[ \, S^{-1} \, ({\frac{\partial}{\partial E}} \, S) - ({\frac{\partial}{\partial E}} \, S^{-1} ) \, S \, \right]_c \\
    &= \,  i \times \frac{\partial}{\partial E} \, {\rm tr} \, \left[ T + T^\dagger \right]_c +  {\rm tr} \, \left[ T^\dagger \,  \overleftrightarrow{\frac{\partial}{\partial E}} \, T \, \right]_c \\
  \end{split}
\end{align}

\noindent The first term is linear in the scattering amplitude, while the second term has a quadratic dependence. Writing them in terms of the phase shift $\delta_E$, we obtain

\begin{align}
  \begin{split}
    \label{eq:landau}
    \frac{1}{4 \, i} \, {\rm tr} \, \left[ S^{-1} \, \overleftrightarrow{\frac{\partial}{\partial E}} \, S \right]_c  &\longleftrightarrow  {\frac{\partial \delta_E}{\partial E}} \\
    \frac{1}{4 } \, \frac{\partial}{\partial E} \, {\rm tr} \, \left[ T + T^\dagger  \right]_c \, &\longleftrightarrow  (1-2 \sin^2\delta_E) \times {\frac{\partial \delta_E}{\partial E}} \\
    \frac{1}{4 \, i} \, {\rm tr} \, \left( T^\dagger \,  \overleftrightarrow{\frac{\partial}{\partial E}} \, T \, \right)_c &\longleftrightarrow   2 \sin^2\delta_E  \times {\frac{\partial \delta_E}{\partial E}}.
  \end{split}
\end{align}

If the scattering amplitude is small, it may be sufficient to retain only the linear term. This corresponds to the approximation

\begin{align}
  \begin{split}
    {Q} &\approx \frac{1}{2} \, {\rm Im} \,  \left[   \int d \phi_2 \, i \mathcal{M} \right].
  \end{split}
\end{align}

\noindent Note that the nature of the trace requires that we consider the same momenta, $\vec{k_i} = \vec{k_i^\prime}$, for the in-coming and out-going states of the matrix element. Hence, within the linear assumption, only the forward-going amplitude is involved in calculating ${Q}$. This is generally not the case when higher order terms are considered.

Before ending this section, we make one further remark on the advantage of writing the phase shift function as ${\rm tr} \ln S$. This formulation makes the generalization to multiple channels intuitive. If in addition to $ 1 + 2 \rightarrow 1 + 2 $, processes such as $ 1 + 2 \rightarrow 3 + 4 $ are also possible. Assuming the simplifications made in Eq.~$(\star)$ are valid, the phase space integrated S-matrix in this case is promoted to a matrix in the reaction channel space. It can expressed in terms of two phase shifts ($\delta_I$, $\delta_{II}$) and an inelasticity parameter $\eta$ as~\cite{K-matrix}

\begin{align}
  S = \left( \begin{array}{cc}
    \eta \, e^{2 \, i \, \delta_I}  &   i \sqrt{1-\eta^2} \, e^{i \, (\delta_I + \delta_{II})} \\
    i \sqrt{1-\eta^2} \, e^{i \, (\delta_I + \delta_{II})}   &  \eta \, e^{2 \, i \, \delta_{II}}
  \end{array} \right).
\end{align}

\noindent Noting the fact that

\begin{align}
    {\rm tr} \ln \, S \rightarrow \ln \, {\rm det} \,[S], 
\end{align}

\noindent the generalization for $Q$ in Eq.~\eqref{eq:Q_def} reads

\begin{align}
  {Q} \rightarrow \delta_I + \delta_{II},
\end{align}

\noindent i.e. it is simply given by the sum of eigenphases~\cite{schott}.

\section{Applications}
\label{sec:3}

\subsection{resonance dominance model}

We first investigate how the HRG model emerges naturally from the S-matrix framework. Assuming resonance production dominates the thermodynamics, we consider the following class of N-body scattering amplitudes describing an s-channel exchange of a resonance:

\begin{align}
	\begin{split}
    i \mathcal{M} &= \frac{-i \, \left\vert \Gamma \right\vert^2 }{M^2 - {\bar{m}_{\rm res}}^2 + i M \gamma} \\
		\gamma &= \frac{1}{2 M} \int d \phi_N \, \vert \Gamma \vert^2,
	\end{split}
\end{align}

\begin{figure*}[ht!]
\centering
\includegraphics[width=0.496\textwidth]{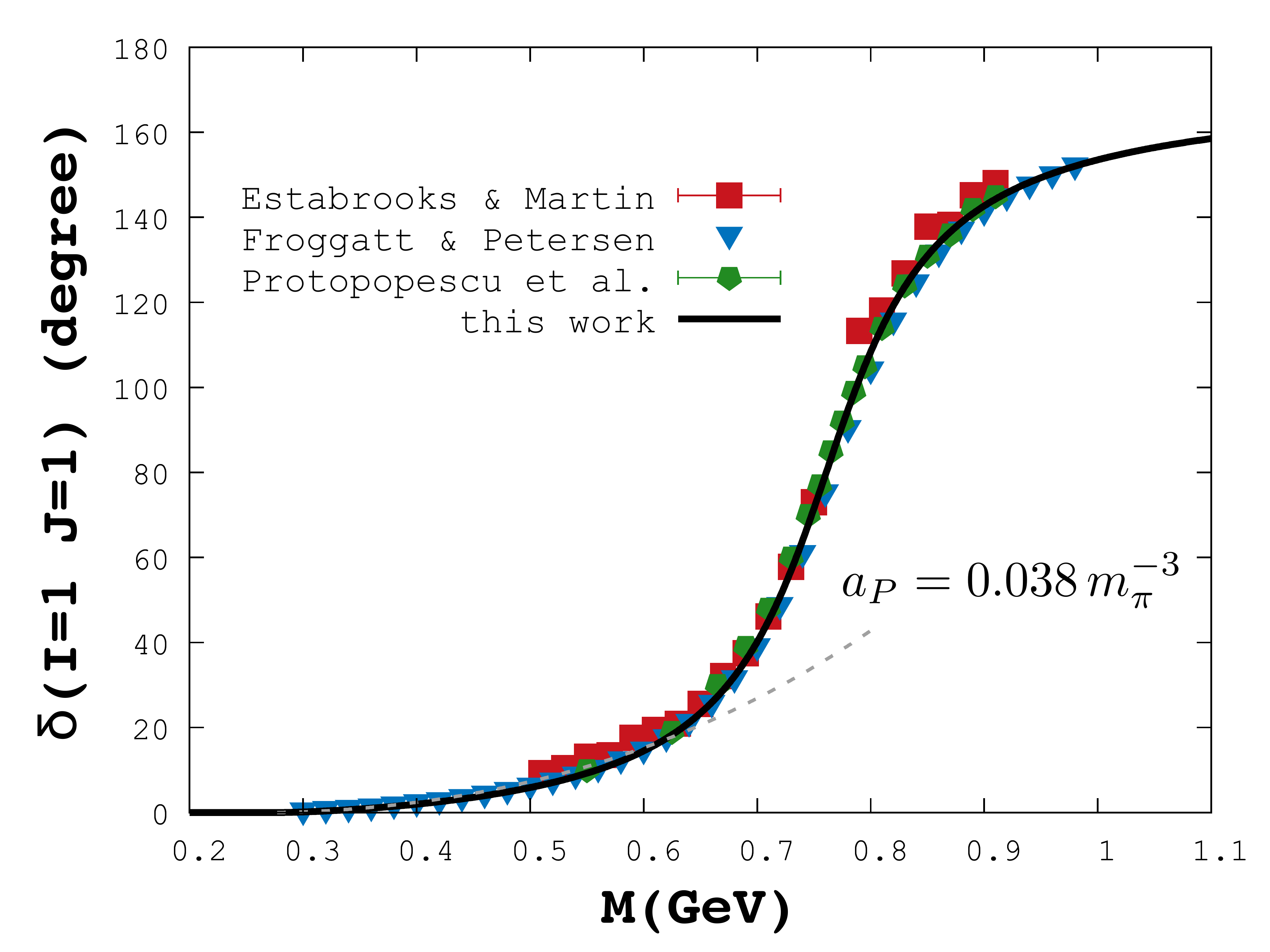}
\includegraphics[width=0.496\textwidth]{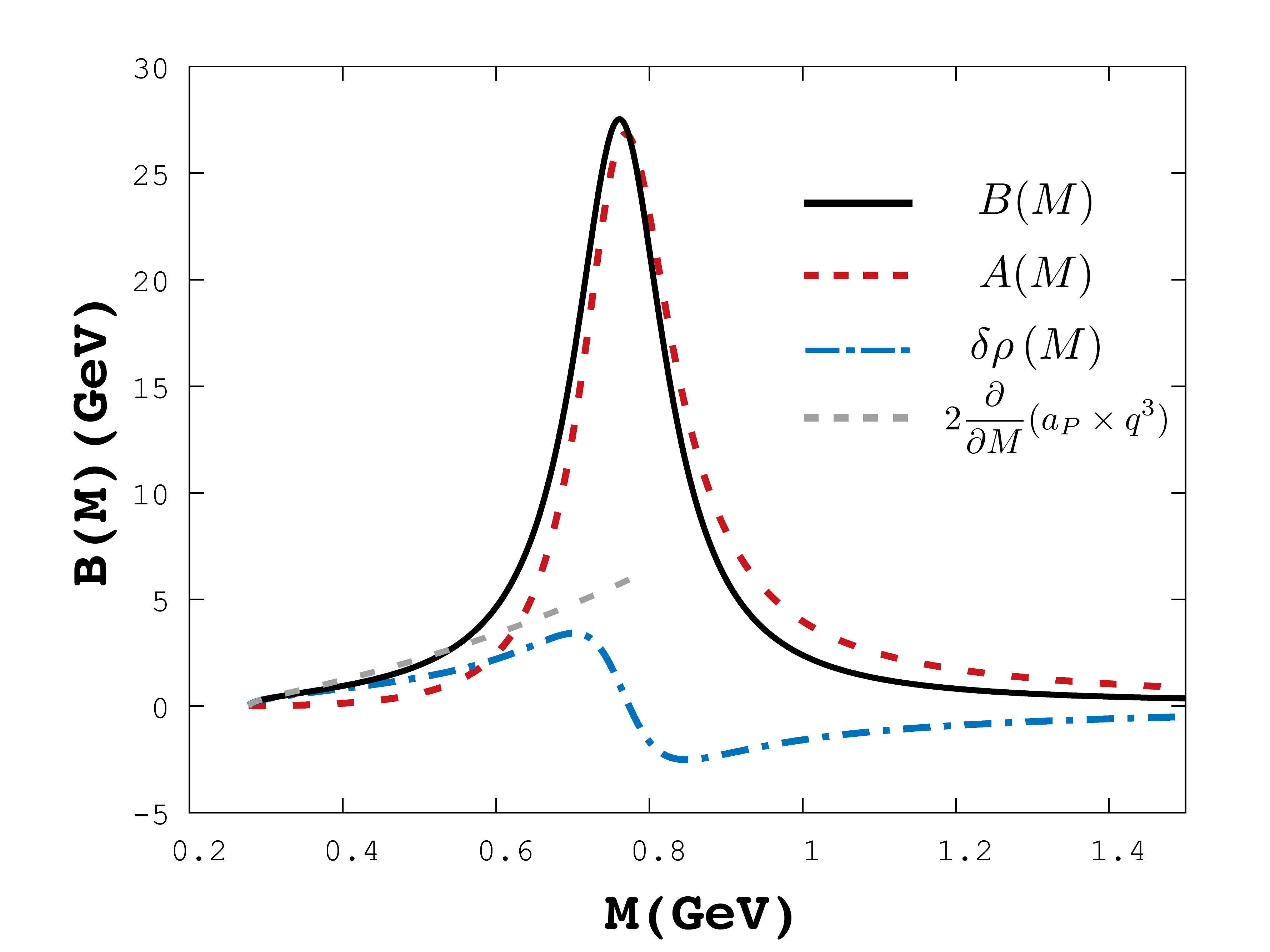}
  \caption{Empirical phase shift data~\cite{Protopopescu:1973sh, Estabrooks:1974vu,Froggatt:1977hu} and the corresponding effective spectral functions for the $\rho$-meson~\cite{rho}.}
\label{fig:one}
\end{figure*}

\noindent where $M$ is the invariant mass of the N-body system, $\bar{m}_{\rm res}$ is the pole mass of resonances. $\phi_N$ denotes the N-body Lorentz invariant phase space, with the explicit expression

\begin{align}
	\begin{split}
    \phi_N  =& \int d \phi_N \\
    =& \int \frac{d^3 p_1}{(2 \pi)^3} \frac{1}{2 E_1}\frac{d^3 p_2}{(2 \pi)^3} \frac{1}{2 E_2}\cdots \frac{d^3 p_N}{(2 \pi)^3}\frac{1}{2 E_N} \times  \\
    &  \hspace{0.25cm} (2 \pi)^4 \, \delta^4(P-\sum_i p_i). 
	\end{split}
\end{align}

Given a model of $\Gamma$ or $\gamma$ for the width, the generalized phase shift function ${Q}(M)$ works out to be

\begin{align}
	\begin{split}
		{Q}(M) &= \frac{1}{2} \,  {\rm Im} \, \left[ \ln{( 1 + \int d \phi_N \, i \mathcal{M} )} \right] \\
                  &= \frac{1}{2} \,  {\rm Im} \, \left[ \ln{( 1 +  \frac{- i \, 2 M \gamma  }{M^2 - {\bar{m}_{\rm res}}^2 + i M \gamma}         )} \right] \\
    &= \tan^{-1} \frac{-M \gamma}{M^2- \bar{m}_{\rm res}^2}.
	\end{split}
\end{align}

\noindent The effective spectral function $B(M)$, and the standard spectral function $A(M)$, can be computed as follows~\cite{kappa,Weinhold:1997ig}:

\begin{align}
	\label{eq:B}
    B(M) &= 2 \, {\frac{\partial}{\partial M}} {Q}(M)
\end{align}

\begin{align}
	\label{eq:A}
  A(M) &= -2 M \, \frac{\sin{2 {Q}(M)}}{M^2-\bar{m}_{\rm res}^2}.
\end{align}

In case of a very narrow resonance, i.e., the limit of $\gamma \rightarrow 0$, we observe that the phase shift function behaves like a theta-function 

\begin{align}
  {Q}(M)   \rightarrow \pi \times \theta(M-\bar{m}_{\rm res}),
\end{align}

\noindent and the two spectral functions converge to same limit:

\begin{align}
  B(M) \approx A(M) \rightarrow 2 \, \pi \times  \delta(M-\bar{m}_{\rm res}).
\end{align}

\noindent This establishes the fundamental premise of the HRG model~\cite{Hagedorn:1965st,Arriola:2014bfa}: contribution of resonances to the thermodynamics is given by an uncorrelated gas of zero-width particles.

On the other hand, Eq.~\eqref{eq:B} is applicable even for a broad resonance. It has been pointed out by Weinhold {\it et al.}~\cite{Weinhold:1997ig} that the effective spectral function $B(M)$ contains both the contribution from the full spectral function $A(M)$ of the resonance and a non-resonant contribution $\delta\rho \, (M)$ from the correlated pair of the forming constituents~\footnote{The separation of $B(M)$ into the two pieces, however, is model dependent. Also it is a different separation from the one presented in Eq.~\eqref{eq:landau}} :

\begin{align}
  B(M) = A(M) + \delta\rho \,(M).
\end{align}

As an example, we consider the effective spectral function of a $\rho$-meson based on the experimental phase shift data~\cite{Protopopescu:1973sh,Estabrooks:1974vu,Froggatt:1977hu}. 
The parametrization of the phase shift function employed in fitting the data is discussed in Ref.~\cite{rho} and will not be repeated here. The phase shift and the spectral functions derived from it via Eq.~\eqref{eq:B}-~\eqref{eq:A} are shown in Fig.~\ref{fig:one}. 

Focusing on the behavior of the spectral functions. Near the pole mass of the resonance, it is seen that $A(M)$ dominates the contribution to the effective spectral function $B(M)$. The net effect of $\delta\rho \, (M)$ is to shift the spectral function towards the lower mass region. Nevertheless, it should be noted that near the threshold $\delta\rho \,(M)$ dominates. In fact, the behavior of $B(M)$ at the threshold is uniquely specified by the chiral symmetry via the scattering length (see Sec.~\ref{sec:non-res}).

In the most commonly used statistical models the width of resonances is sometimes implemented via a Breit-Wigner function, on the other hand, the effect of $\delta\rho \,(M)$ is mostly neglected. The importance of the latter term depends on the observable under study. For the $p_T$-spectra of the decay pions coming from $\rho$-mesons, it is found that the non-resonant term can contribute substantially to the soft part of the $p_T$-spectrum~\cite{rho}.

\subsection{Non-resonant scattering at threshold}
\label{sec:non-res}

Not all interaction channels among hadrons are resonance dominated. Some are purely repulsive and some have complicated energy dependence due to the intricate hadronic interaction. The S-matrix formalism presented here can consistently take these into account~\cite{exvol}.

Near the threshold, when the momenta of the scattering particles are small, the interaction can be reliably described within an effective field theory framework~\cite{Kaplan:1998tg} or even by quantum mechanical models. The phase shift, to lowest order in the momentum, takes the following form

\begin{align}
    {Q}(M) &\approx d_S \times (a_S \, q) + d_P \times (a_P \, q^3) + \cdots,
\end{align}

\noindent where $q$ is the momentum of the particles in the center of mass frame

\begin{align}
  q = \frac{1}{2} \, M \, \sqrt{1 + \frac{(m_1+m_2)^2}{M^2}} \, \sqrt{1 + \frac{(m_1-m_2)^2}{M^2}}.
\end{align}

For a concrete example, consider the case of $\pi \pi$ scattering, we get

\begin{align}
    {Q}(M) \approx (a_S^{I=0} + 5 \, a_S^{I=2}) \times q + 9 \, a_P^{I=1} \times q^3.
\end{align}

\noindent In this case, the scattering lengths are well constrained by the chiral perturbation theory. Moreover, it has been noted that there is an essential cancellation effect~\cite{Venugopalan:1992hy,sigma} between the $I = 0$ and the $I = 2$ channel in the low invariant mass region. This results in a very small S-wave contribution to the thermodynamics. Similar conclusion applies to the $\kappa$-meson~\cite{kappa}.

\subsection{Structureless N-body scattering}

A particularly simple case in which the exact N-body generalized phase shift function ${Q}_N(M)$ can be readily extracted is the model of structureless scattering.
Assuming the general N-body scattering matrix is to be described by a dimensionful ($\sim E^{2N-4}$) coupling constant $\lambda_N$ such that

\begin{align}
	\label{eq:structureless}
	\begin{split}
		i \mathcal{M} &= i \, \lambda_N,
	\end{split}
\end{align}

\noindent with

\begin{align}
	\label{eq:ps}
		{Q}_N(M) = \frac{1}{2} \, {\rm Im} \, \left[ \ln{( 1 + i \, \lambda_N  \times \phi_N)} \right].
\end{align}

\noindent The problem of calculating the phase shift function then boils down to the determination of the N-body phase space function $\phi_N(M)$. An efficient way to accomplish this task is to employ the K{\"a}ll{\'e}n expansion~\cite{Byckling:1971vca}, which provides a recursive definition of the N-body phase space function $\phi_N(s=M^2)$, via

\begin{align}
  \begin{split}
  \phi_N(s) = \frac{1}{16 \, \pi^2 s} \, \int_{s^\prime_-}^{s^\prime_+} d s^\prime  \, \sqrt{\lambda(s,s^\prime,m_N^2)} \, \times \\
  \phi_{N-1}(s^\prime, m_1^2, m_2^2, ..., m_{N-1}^2),
  \end{split}
\end{align}

\noindent where $\lambda(x,y,z)$ is the K{\"a}ll{\'e}n triangle function

\begin{align}
\lambda(x,y,z) = x^2 + y^2 + z^2 - 2 x y - 2 x z - 2 y z,
\end{align}

\noindent and

\begin{align}
  \begin{split}
  s^\prime_+ &= (\sqrt{s}-m_N)^2 \\
  s^\prime_- &= (\sum_{i=1}^{N-1} m_i)^2.
  \end{split}
\end{align}

\noindent For the case of massless particles ($m_i = 0$), the integral can be performed analytically and the exact expression for the N-body phase space reads

\begin{align}
  \begin{split}
  \label{eq:ps_lim}
  \phi^{\rm massless}_N(s) &= a_N \times s^{N-2} \\
  a_N &= 2 \, \pi \times (\frac{1}{16 \, \pi^2})^{N-1} \, \frac{1}{(N-2)!\,(N-1)!}.
  \end{split}
\end{align}

\begin{figure*}[ht!]
	\centering
 \includegraphics[width=0.8\textwidth]{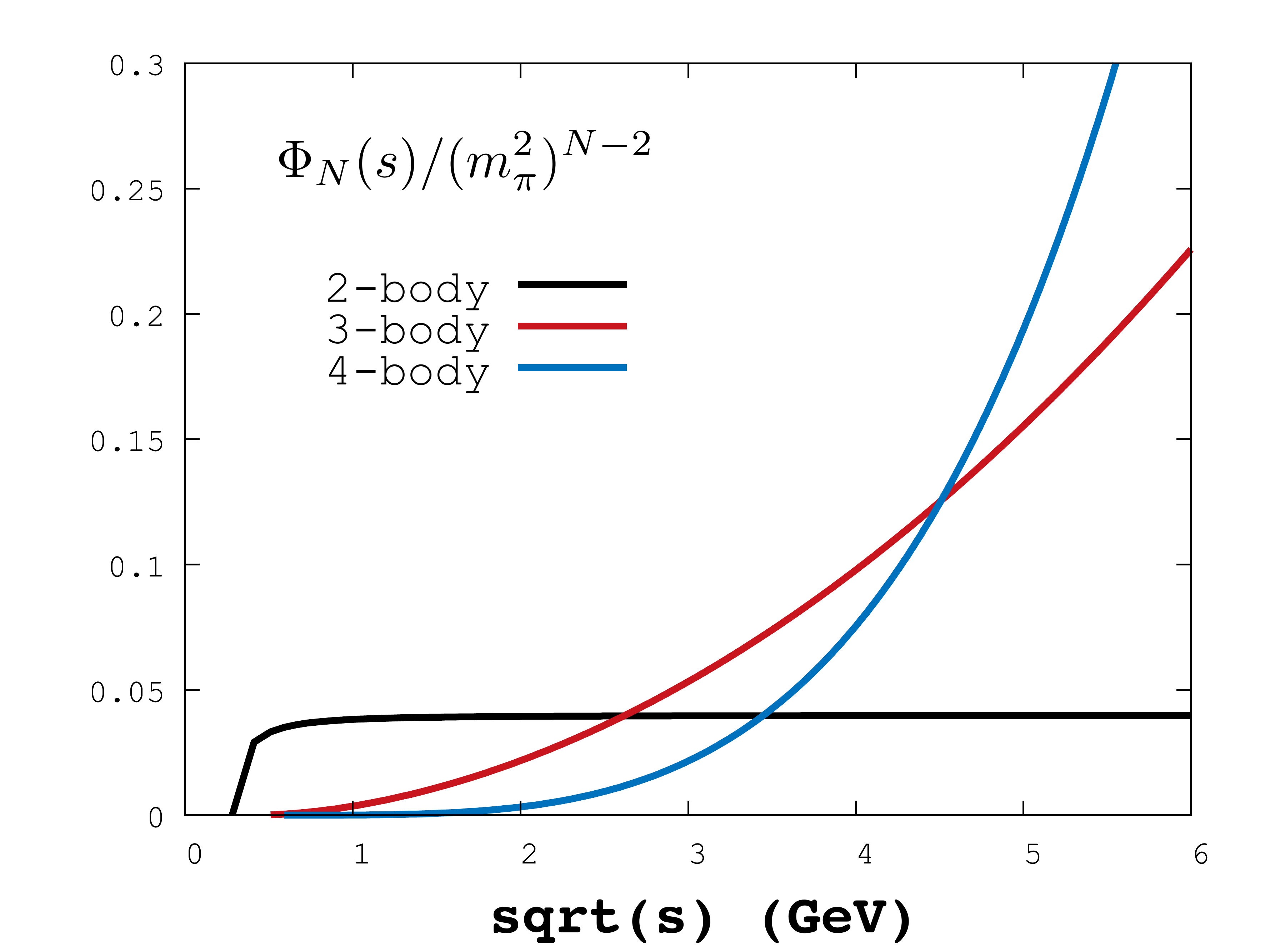}
  \caption{$(N= 2, 3, 4)$-body phase space functions~\eqref{eq:ps}, scaled by the appropriate powers of pion mass, versus the center of mass $\sqrt{s}$. }
	\label{fig:two}
\end{figure*}

To demonstrate how the N-body phase space function increases with $\sqrt{s}=M$, we compute Eq.~\eqref{eq:ps} numerically for a system of pions. For a meaningful comparison, we scale the dimensionful phase space function with the appropriate powers of $m_\pi$. The result is shown in Fig.~\ref{fig:two}.

From a purely kinematical point of view, effects from higher N-body phase space are generally suppressed compared to the lower ones at low invariant mass. However, their effects will show up and will eventually be dominating at high invariant masses in the manner dictated by Eq.~\eqref{eq:ps_lim}. In case of finite density or chemical potential, the takeover by higher N-body phase spaces can occur more rapidly due to the fugacity factor $(e^{\mu/T})^N$ associated with an N-body state. 

The phase space dominance model discussed here may be of interest to phenomenological studies. Performing modeling on the level of S-matrix elements or amplitudes, e.g. the invariant mass dependence of $\lambda_N$, can establish closer connection between observables and model parameters. Moreover, symmetries and physical conditions can be imposed on the S-matrix elements to constrain their functional form. 
On passing, we note that similar models~\cite{Hormuzdiar:2000vq, Rischke:2001bt, Koch:2002uq, Becattini:2004td} have been applied to investigate the emergence of thermal-like behavior of the particle spectra at freezeout conditions.

In the next section~\ref{sec:4}, we investigate how the purely kinematical consideration presented here is modified by interaction dynamics in some simple cases of 3- and 4-body scattering.

\section{Further examples}
\label{sec:4}

\subsection{$3$-body process: the triangle diagram}

\begin{figure}[ht!]
	\centering
 \includegraphics[width=0.45\textwidth]{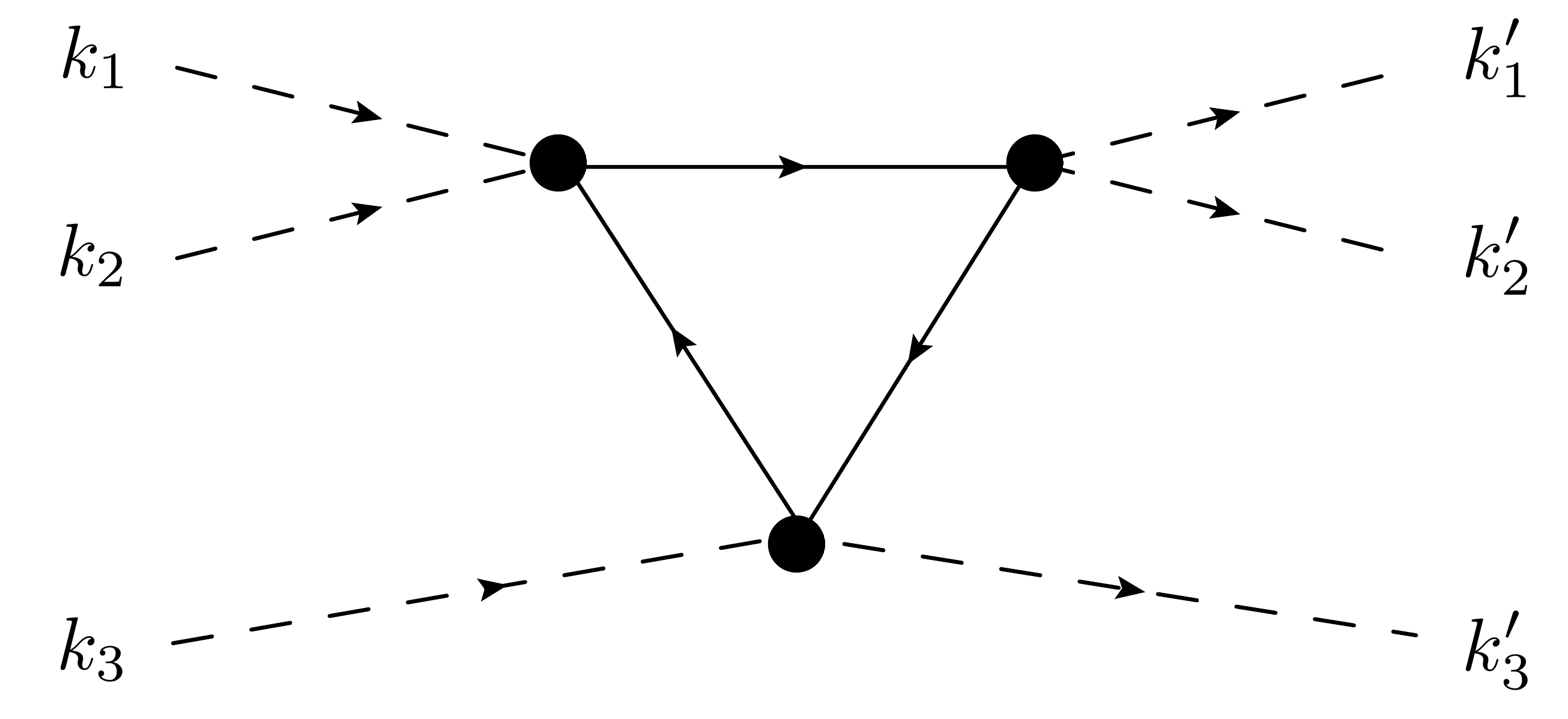}
  \caption{Set up of the Triangle diagram.}
 \label{fig:three}
\end{figure}


The triangle diagram is usually studied in a quantum field theory with a $\lambda_3 \phi^3$ interaction.
Here we consider an alternative scenario where a fully connected 3-body amplitude is dynamically generated by 2-body scatterings, i.e. particles interact two at a time, forming an effective triangle diagram as shown in Fig.~\ref{fig:three}~\footnote{ 
We pay special attention to this diagram for the following reason. In a classical system of particles interacting via a pair-wise potential, it can be shown~\cite{Pathria} that only Mayer graphs with such a closed triangle topology enters the calculation of the third virial coefficient. Those constructed with two links (open triangle), though present in the cluster expansion, are absent in the virial expansion. See also Ref.~\cite{Vasiliev:1998cq} on the definition of one-vertex-irreducible (1VI) graphs.}:

For our purpose we only consider the lowest order term in $\lambda$ for the topology of the diagram of interest. First, we write down the amplitude of the process according to Feynman rules:

\begin{align}
  \begin{split}
    i \mathcal{M}^{ \bigtriangleup }(q_1, q_2, q_3) =& \int \frac{d^4 l}{(2 \pi)^4} \, (-i \, \lambda)^3 \, \times   \\
    &  i \, G(l) \times i \, G(l+q_1) \times  i \, G(l-q_2) 
	\end{split}
\end{align}

\begin{figure*}[ht!]
	\centering
 \includegraphics[width=0.48\textwidth]{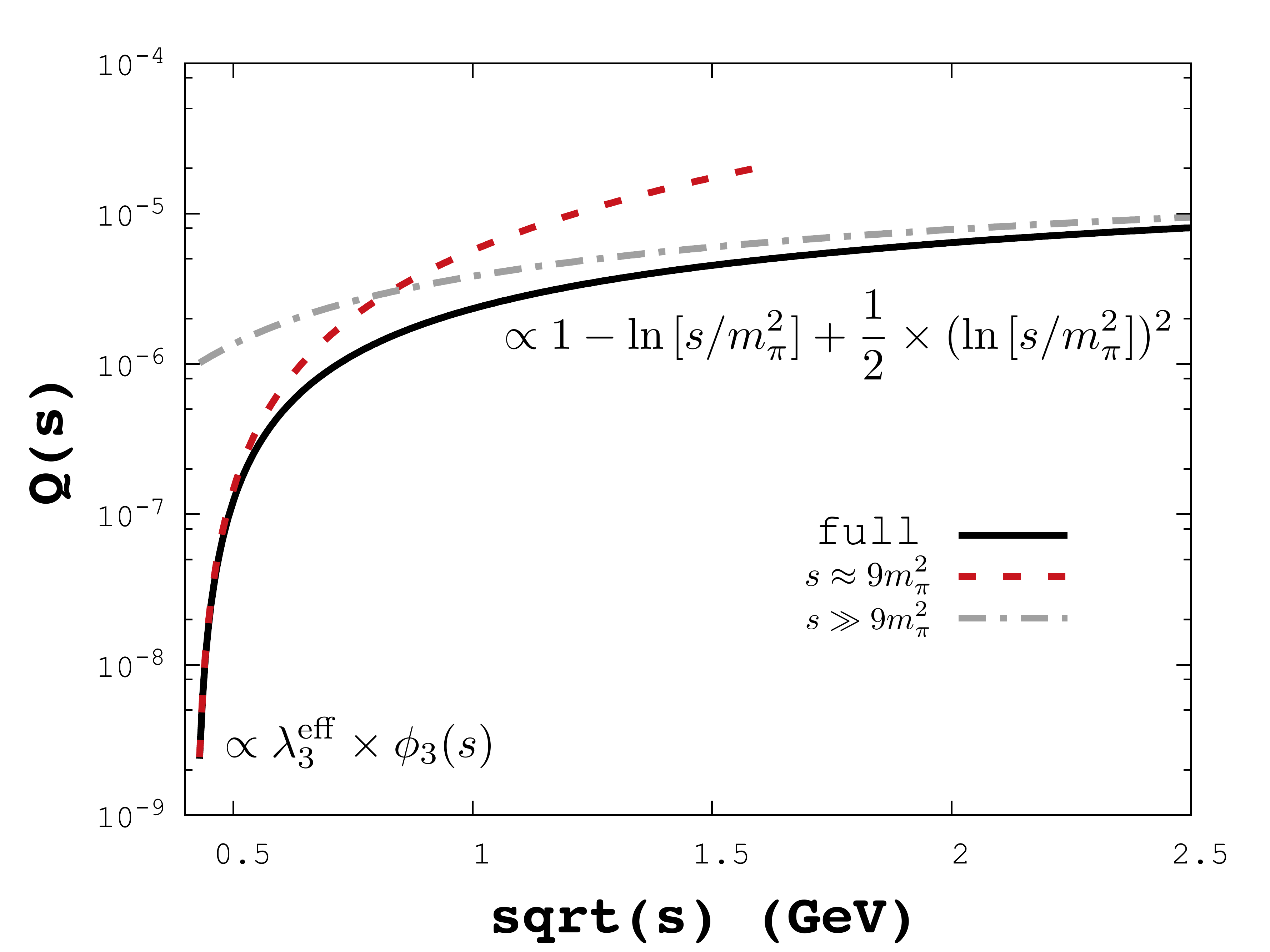}
 \includegraphics[width=0.48\textwidth]{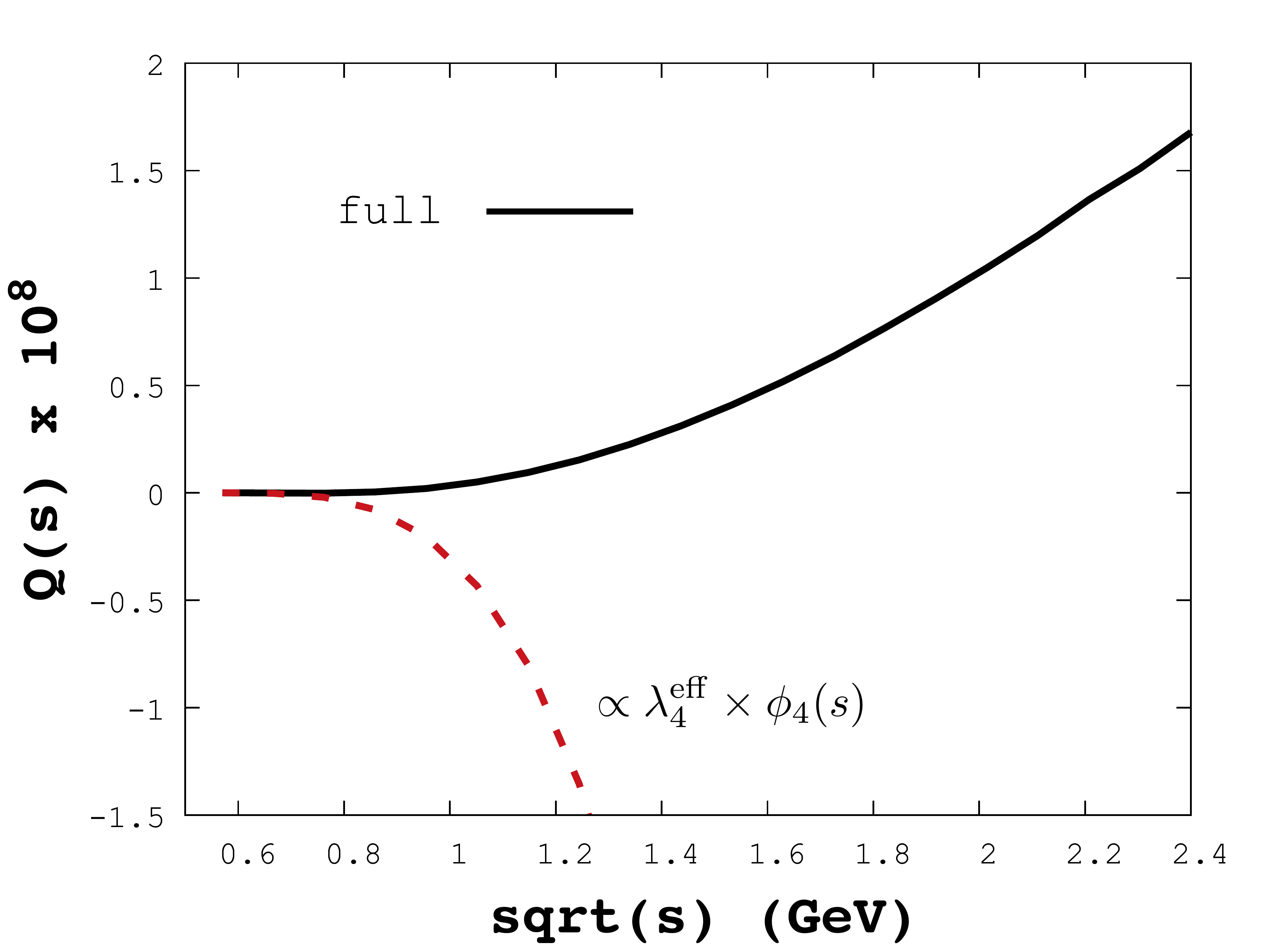}
  \caption{The generalized phase space function ${Q}(s)$ for triangle (left) and box (right) diagrams under an on-shell condition discussed in the text.}
 \label{fig:four}
\end{figure*}

\noindent where

\begin{align}
	\begin{split}
    G(k) &= \frac{1}{k^2-m_\pi^2+ i\epsilon} \\
    q_1 &= k_1 + k_2 \\
    q_2 &= k_3 - k_3^\prime \\
    q_3 &= k_1^\prime + k_2^\prime = q_1 + q_2 \\
    P_I &= k_1 + k_2 + k_3 = k_1^\prime +  k_2^\prime +  k_3^\prime.
	\end{split}
\end{align}

\noindent Using the standard Feynman's trick and computing the loop momentum integral using dimensional regularization, we obtain

\begin{align}
    i \mathcal{M}^{ \bigtriangleup }(q_1^2, q_2^2, s=P_I^2) &= -i \frac{ \lambda^3}{16 \pi^2} \, \int_0^1 d x \, \int_0^{1-x} d y \,  \frac{1}{\Delta(x,y)} 
\end{align}

\begin{align}
  \begin{split}
    \Delta(x,y) &= m_\pi^2 - x(1-x) \, q_1^2 - y(1-y) \, q_2^2 \\
    &\hspace{0.4cm}-2 \,  x \,  y \, q_1 \cdot q_2 -i \epsilon. 
	\end{split}
\end{align}

\noindent This matrix element does not satisfy the factorization condition $(\star)$ and the general computation of ${Q}$ becomes notoriously difficult. However, if we consider only the linear term in the expansion

\begin{align}
  \label{eq:tri}
    {Q}(s) \approx \frac{1}{2} \,  {\rm Im} \, \left[ \int d \phi_3 \, i {\mathcal{M}}^{\bigtriangleup} \right],
\end{align}

\noindent the integral only involves the amplitude with the following on-shell condition:

\begin{align}
  k^\prime_i = k_i,
\end{align}

\noindent or equivalently

\begin{align}
	\begin{split}
    q_1 &= k_1 + k_2  \\
    q_2 &= 0 \\
    q_3 &= q_1. 
	\end{split}
\end{align}

\noindent The Feynman amplitude in this case can be computed analytically to give

\begin{align}
	\begin{split}
    i \mathcal{M}^{\bigtriangleup, o.s.}(q_1^2, s) &= -i \frac{ \lambda^3}{16 \, \pi^2} \, \frac{z}{q_1^2} \, \ln \frac{1-z}{1+z} \\
    z &= \frac{1}{\sqrt{1- \frac{4 m_\pi^2}{q_1^2}}}. 
	\end{split}
\end{align}

\noindent The variable $z$ here should be understood to possess a small and negative imaginary part. The $q_1^2$-dependence in the amplitude is crucial since it is one of the integration variables in the 3-body phase space. In fact, writing $s^\prime = q_1^2$, we have in this case

\begin{align}
	\begin{split}
    \int d \phi_3 \, \rightarrow& \,  \frac{1}{128 \pi^3} \frac{1}{s^2} \int_{4 m_\pi^2}^{(\sqrt{s}-m_\pi)^2} d s^\prime \, \times \\
    &\sqrt{\lambda(s,s^\prime,m_\pi^2)} \times \sqrt{\lambda(s^\prime,m_\pi^2, m_\pi^2)}.
	\end{split}
\end{align}

\noindent The numerical result of Eq.~\eqref{eq:tri} is shown in Fig.~\ref{fig:four}.

The behavior of the generalized phase shift function ${Q}(s)$ for the triangle diagram can be qualitatively understood as follows. First, we note that close to the threshold $s \rightarrow 9 \, m_\pi^2$, 

\begin{align}
  \begin{split}
    {\rm Im} \, \left( i \mathcal{M}^{\bigtriangleup, o.s.}(q_1^2, s) \right) &\approx \lambda_3^{\rm eff} \\
     &= \frac{\lambda^3}{16 \pi^2} \frac{1}{2 m_\pi^2},
  \end{split}
\end{align}

\noindent meaning that the function ${Q}$ is simply dictated by the 3-body phase space

\begin{align}
  {Q}(s) &\approx \frac{1}{2} \times \lambda_3^{\rm eff} \times \phi_3(s).
\end{align}

\noindent However, this is no longer the case at large invariant masses. In fact, the following asymptotic expression can be obtained for the phase shift function at $s \gg m_\pi^2$: 

\begin{align}
  \begin{split}
    {Q}(s) &\approx \frac{\lambda^3}{8192 \, \pi^5} \, \int_{\xi_0}^1 d \xi \, (\frac{1}{\xi}-1) \, \left[-z \, \ln \left\vert \frac{1-z}{1+z} \right\vert\right] \\
  &\approx \frac{\lambda^3}{4096 \, \pi^5} \times \left[ 1 + \ln \frac{\xi_0}{4} + ( \ln \frac{\xi_0}{4} )^2 \right] \\
  \end{split}
\end{align}

\noindent where

\begin{align}
  \begin{split}
    z &= \frac{1}{\sqrt{1-\frac{\xi_0}{\xi}}} \\
    \xi_0 &= \frac{4 m_\pi^2}{s}.
  \end{split}
\end{align}

We find that the structureless scattering approximation is valid only for a very narrow invariant mass range near the threshold. The phase shift function tends to be more suppressed when the dynamics is taken into account. The approach to the expected asymptotic limit at $s \gg  m_\pi^2$ is rather slow due to the presence of logarithmic terms. This may be interesting for phenomenological studies and reaction simulations when modeling the space-time details of an N-body scattering.

Nevertheless, since only the lowest order diagram (in coupling $\lambda$) with triangle topology is considered, the invariant mass dependence obtained here is only schematic. More sophisticated method for obtaining the 3-body T-matrix, e.g. by solving the Faddeev equation, and the inclusion of the non-linear terms are needed to produce a realistic assessment of its effect in thermodynamics.

\subsection{$4$-body process: the box diagram}

\begin{figure}[h]
	\centering
 \includegraphics[width=0.45\textwidth]{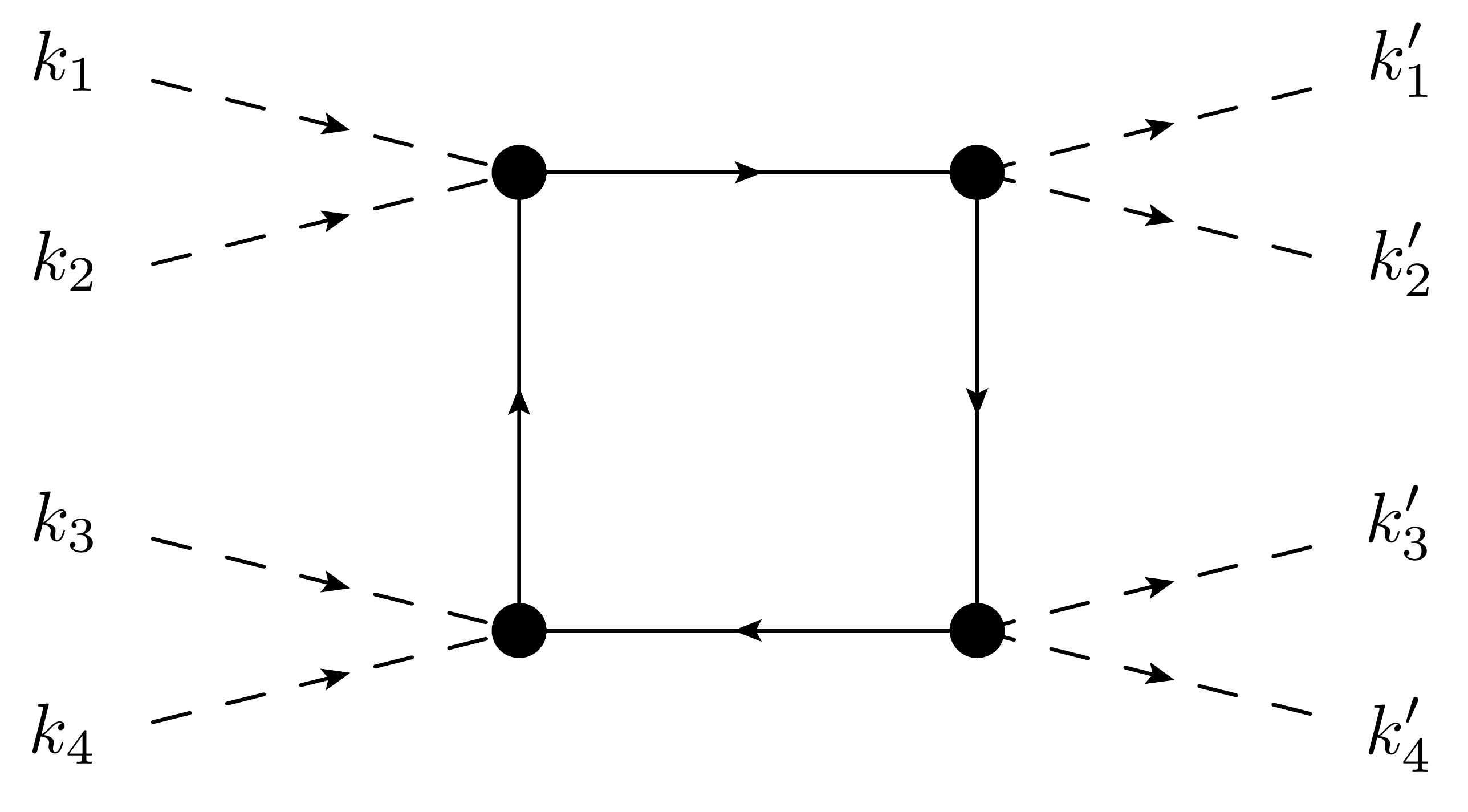}
  \caption{Set up of the Box diagram.}
	\label{fig:five}
\end{figure}

An analogous study can be performed for a 4-body scattering. Here we choose the familiar box diagram, again dynamically generated by 2-body scatterings, see Fig.~\ref{fig:five}.

\begin{align}
	\begin{split}
    i \mathcal{M}^{ \rm box }(q_1, q_2, q_3, q_4) &= \int \frac{d^4 l}{(2 \pi)^4} \, (-i \, \lambda)^4 \times \\
    &i \, G(l) \times i \, G(l+q_1) \times  \\
    &i \, G(l+q_1-q_3) \times i \, G(l-q_2) \\
	\end{split}
\end{align}

\noindent with

\begin{align}
	\begin{split}
    G(k) &= \frac{1}{k^2-m_\pi^2+ i\epsilon} \\
    q_1 &= k_1 + k_2 \\
    q_2 &= k_3 + k_4 \\
    q_3 &= k_2^\prime + k_2^\prime \\
    q_4 &= k_3^\prime + k_4^\prime \\
    P_I &= k_1 + k_2 + k_3 + k_4 = k_1^\prime +  k_2^\prime +  k_3^\prime + k_4^\prime.
	\end{split}
\end{align}

Using the standard Feynman's trick and computing the loop momentum integral using dimensional regularization:

\begin{align}
	\begin{split}
    \label{eq:box}
   i \mathcal{M}^{\rm box}(q_1, q_2, q_3, q_4) &= i \frac{ \lambda^4}{16 \pi^2} \, \int_0^1 d x \, \int_0^{1-x} d y \, \times \\ 
    &\hspace{0.2cm} \int_0^{1-x-y} d z \, \times \left(\frac{1}{\Delta(x,y,z)} \right)^2 \\
	\end{split}
\end{align}

\begin{align}
	\begin{split}
    \Delta(x,y,z) &= m_\pi^2 -  \{ x(1-x) \, q_1^2 \\
    &\hspace{0.2cm}+ y(1-y) \, (q_1-q_3)^2 + z(1-z) \, q_2^2 \\
    &\hspace{0.2cm} - 2 \, x \, y \, q_1 \cdot (q_1-q_3) + 2 \, x \, z \, q_1 \cdot q_2  \\
    &\hspace{0.2cm}+ 2 \, y \, z \, q_2 \cdot (q_1-q_3 ) -i \epsilon\}. 
	\end{split}
\end{align}

The full matrix element does not satisfy the factorization condition $(\star)$ and the general computation of ${Q}$ is vastly complicated. Considering only the linear term,

\begin{align}
    {Q}(s) \approx \frac{1}{2} \,  {\rm Im} \, \left[ \int d \phi_4 \, i {\mathcal{M}}^{\rm box, o.s.} \right],
\end{align}

\noindent allows us to focus on the following simplified on-shell condition:

\begin{align}
  k^\prime_i = k_i,
\end{align}

\noindent which gives 

\begin{align}
	\begin{split}
    q_1 &= k_1 + k_2 \\
    q_2 &= k_3 + k_4 \\
    q_3 &= q_1 \\
    q_4 &= q_2. 
	\end{split}
\end{align}

\noindent Even with this simplified on-shell condition, it is not clear whether the integral~\eqref{eq:box} over the Feynman parameters can be computed in closed form. Nevertheless, numerical computation of the real and imaginary parts can be robustly performed. 

The dependence on $q_1^2$ and $q_2^2$ of the amplitude $\mathcal{M}^{\rm box, o.s.}$ is crucial in correctly determining the phase shift function ${Q}$. Here, the numerical integration over the 4-body phase space is much more involved. However, for amplitudes depending only on $s_1^\prime = q_1^2$ and $s_2^\prime = q_2^2$, the following method of integrating the phase space can be applied:

\begin{align}
	\begin{split}
    \int d \phi_4 \, \rightarrow&   \frac{1}{4 \pi^2} \int_{4 m_\pi^2}^{(\sqrt{s}-2 m_\pi)^2} d s_1^\prime \,  \int_{4 m_\pi^2}^{(\sqrt{s}-2 m_\pi)^2} \, d s_2^\prime \times \\
    & \phi_2(s_1^\prime, m_\pi^2, m_\pi^2) \times \phi_2(s_2^\prime, m_\pi^2,m_\pi^2) \times \\
    &\times \phi_2(s, s_1^\prime, s_2^\prime) \times \theta( \sqrt{s}-\sqrt{s_1^\prime}-\sqrt{s_2^\prime} ).
	\end{split}
\end{align}

\noindent This reduces the original 8-dimensional integral into a 2-dimensional one with the suitable integrating variables for $\mathcal{M}^{\rm box, o.s.}$. The numerical result is shown in Fig.~\ref{fig:four}. 

As in the triangle diagram case, we study the expected behavior of the phase shift function, near the 4-body threshold. We find

\begin{align}
  \begin{split}
    {\rm Im} \, \left( i \mathcal{M}^{\rm box, o.s.}(q_1^2, q_2^2, s) \right) &\approx  \lambda_4^{\rm eff} \\
    &= \frac{\lambda^4}{256 \pi^2} \frac{1}{m_\pi^4} \times (\frac{\sqrt{3}}{2} \, \ln \, (7-4 \, \sqrt{3}) + 2), 
  \end{split}
\end{align}

\noindent which gives the following ${Q}(s)$ near the threshold

\begin{align}
  {Q}(s) &\approx \frac{1}{2} \times \lambda_4^{\rm eff} \times \phi_4(s).
\end{align}

\noindent The numerical value of the effective constant turns out to be negative, and as seen in Fig.~\ref{fig:four}, the full result deviates from the threshold behavior very rapidly as $s$ increases.

It should be noted that the simplified examples considered here are by no means capturing the full complexity of the full N-body amplitude.
Nevertheless, when a realistic amplitude is supplied by a model, the scheme can be directly employed to assess its contribution to the thermodynamics.

\subsection{towards a cluster/virial expansion using the S-matrix approach}

Recall the cluster expansion for the logarithm of the grand partition function 

\begin{align}
  \label{eq:p_clus}
  \frac{P(\xi)}{k_B T} = \frac{\ln Z}{V} = n_0 \, \sum_l b_l \, \xi^l
\end{align}

\noindent where $n_0$ is some density scale usually taken to be $1/\lambda^3$ with $\lambda$ being the thermal wavelength of the particle. The expansion is in powers of the fugacity $\xi$, and the dimensionless coefficients $b_l$'s are related to the cluster integral. The density can be obtained via

\begin{align}
  \label{eq:den_clus}
  n = \xi \frac{\partial}{\partial \xi} \frac{P(\xi)}{k_B T} = n_0 \sum_l l \times b_l \, \xi^l.
\end{align}

\noindent The virial expansion can be obtained by re-expressing the pressure $P$ in terms of density $n$:

\begin{align}
  \label{eq:virial}
  \frac{P}{n \, k_B T} = \frac{\sum b_l \, \xi^l}{\sum l \times b_l \, \xi^l} = \sum_l a_l \times (\frac{n}{n_0})^l.
\end{align}

\noindent Here $a_l$'s are the dimensionless virial coefficient, and are related to $b_l$'s by

\begin{align}
  \begin{split}
    a_1 &= 1  \\
    a_2 &= -b_2  \\
    a_3 &= -2 b_3  + 4 b_2^2 \\
    \cdots.
  \end{split}
\end{align}

The S-matrix expansion in Eq.~\eqref{eq:main} naturally lends itself to the form of a cluster expansion. In fact, the coefficients $b_l$'s are related to the connected $l$-body S-matrix element. In particular, for the 2-body case, one obtains the interaction part of $b_2$ by

\begin{align}
  \Delta b_2 = \frac{1}{n_0} \times \int \frac{d^3 P}{(2 \pi)^3} \,  \frac{ d M}{2 \pi} \, e^{- \beta \sqrt{P^2 + M^2}} \, 2 \, {\frac{\partial}{\partial M}} {Q}_2,
\end{align}

\noindent and similarly for higher $b_l$'s. 

While the classical cluster coefficients $b_l$'s are well known~\cite{Pathria} and can be computed readily once the potential between the particles is given, the quantum version of $b_l$'s beyond the 2-body case are much more challenging to obtain. Nevertheless, important progress has been made for the system of ultracold Fermi gas~\cite{Bedaque:2002xy, Leyronas, Kaplan:2011br, Liu}.

Evaluating the higher virial terms is an extremely important task. In addition to checking the validity of the common implementation of retaining only the 2-body term, i.e., the Beth-Uhlenbeck formula, it may also help in extending the applicability of the S-matrix formalism to study the equation of state for a dense system. This may seem implausible since the virial expansion is essentially an expansion in density, and at high density it is destined to diverge. Nevertheless, even in this situation studying the radius of convergence may reveal important physics of the medium such as the existence of a critical point. Moreover, for some specific systems like the classical lattice gas, it is possible to constructs a high-density expansion~\cite{Lee:1952ig,Ushcats1,Ushcats2} (effectively an expansion in $\xi^{-1}$) for the thermodynamic potential. The coefficients in such an expansion are found to be related to the $b_l$'s in the standard cluster expansion. In any case, it would be beneficial to gain a solid knowledge of the higher virial terms.

The current treatment of the S-matrix expansion involving the separation of kinematics (N-body phase space) and dynamics (amplitude) may contribute to clarifying these issues. In particular, we notice that the virial expansion can be understood as a kind of Legendre transform of the cluster expansion. Hence, the S-matrix diagrams that are involved in the virial expansion is expected to be of the skeleton type (or one vertex irreducible~\cite{Vasiliev:1998cq}). This may help to shorten the list of diagrams in the calculation. Research along this direction is underway.

\section{Conclusions}

We have applied a phase space expansion scheme to evaluate the N-body trace in the S-matrix formulation of statistical mechanics.
A generalized phase shift function, suitable for studying the thermodynamical contribution of $N \rightarrow N$ processes, is proposed and explicitly worked out in some simple cases. Using the expansion scheme we have revisited how the hadron resonance gas (HRG) model emerges from the S-matrix framework, and how resonance widths can be consistently included, together with the non-resonant interactions. Extension to the general N-body cases within the phase space dominance scenario is worked out, and the influence of dynamics within some simple models are studied. 

The framework presented here is flexible enough to encompass many theoretical approaches. The key input is an N-body S-matrix element, which can be obtained from the experiment (e.g. scattering phase shifts), or be calculated within models. In this work, we have focused on the connection to field theoretical models via the quantum amplitudes. In fact, the S-matrix elements can as well be obtained from potential models~\cite{Barnes:1991em, Lacroix:2014gsa} or from lattice calculations~\cite{Briceno:2017tce}. Moreover, from a phenomenological point of view, it may be more intuitive to perform modeling at the level of S-matrix elements since these are more connected to the experimentally measured quantities. Furthermore, symmetries and other physical conditions may be imposed on the S-matrix elements to constrain their functional form in the space of model parameters.

Much of the discussions presented here builds on the idea that the quantity $d \delta/d E$ represents a change of the density of state due to the interaction. An interesting alternative interpretation of this quantity, given in Ref.~\cite{Danielewicz}, is the concept of time delay: particles spend longer or shorter in the interaction region due to the attractive or repulsive nature of the interaction. This has been applied to prescribing the space-time details of particle scatterings in a reaction simulation. It would be interesting to see how these higher N-body contributions may be systematically included in this context, and more importantly, their effects on heavy ion collision observables like the transport coefficients.

\acknowledgements

I acknowledge fruitful discussions with Eric Swanson and Bengt Friman.
I thank Andre Leclair and Robert Pisarski for giving constructive comments.
I am also grateful to Krzysztof Graczyk for stimulating discussions on N-body amplitudes.
This work was partly supported by the Polish National Science Center (NCN), under Maestro grant DEC-2013/10/A/ST2/00106 and by the Extreme Matter Institute EMMI, GSI.


\begin{thebibliography}{20}

\bibitem{dmb} 
  R.~Dashen, S.~K.~Ma and H.~J.~Bernstein,
  Phys.\ Rev.\  {\bf 187}, 345 (1969).

\bibitem{Venugopalan:1992hy}
  R.~Venugopalan and M.~Prakash,
  Nucl.\ Phys.\ A {\bf 546} (1992) 718.


\bibitem{sigma}
  W.~Broniowski, F.~Giacosa and V.~Begun,
  Phys.\ Rev.\ C {\bf 92} (2015) 034905.

\bibitem{kappa}
  B.~Friman, P.~M.~Lo, M.~Marczenko, K.~Redlich and C.~Sasaki,
  Phys.\ Rev.\ D {\bf 92} (2015) 074003.

\bibitem{Maris:2003vk} 
  P.~Maris and C.~D.~Roberts,
  Int.\ J.\ Mod.\ Phys.\ E {\bf 12}, 297 (2003).

\bibitem{Popovici:2011yz} 
  C.~Popovici, P.~Watson and H.~Reinhardt,
  Phys.\ Rev.\ D {\bf 83}, 125018 (2011).


\bibitem{Beth:1937zz} 
E.~Beth and G.~Uhlenbeck,
Physica {\bf 4}, 915 (1937).

\bibitem{How:2010zz} 
  P.~T.~How and A.~LeClair,
  Nucl.\ Phys.\ B {\bf 824}, 415 (2010).


\bibitem{Weinhold:1997ig}
  W.~Weinhold, B.~Friman and W.~N\"orenberg,
  Phys.\ Lett.\ B {\bf 433}, 236 (1998).



\bibitem{K-matrix} 
  S.~U.~Chung, J.~Brose, R.~Hackmann, E.~Klempt, S.~Spanier and C.~Strassburger,
  Annalen Phys.\  {\bf 4}, 404 (1995).


\bibitem{schott} 
  S.~Pratt, P.~Siemens and Q.~N.~Usmani,
  Phys.\ Lett.\ B {\bf 189}, 1 (1987).

\bibitem{Hagedorn:1965st} 
  R.~Hagedorn,
  Nuovo Cim.\ Suppl.\  {\bf 3}, 147 (1965).


\bibitem{Arriola:2014bfa} 
E.~Ruiz Arriola, L.~L.~Salcedo and E.~Megias,
Acta Phys.\ Polon.\ B {\bf 45}, no. 12, 2407 (2014).




\bibitem{Protopopescu:1973sh}
  S.~D.~Protopopescu {\it et al.},
  Phys.\ Rev.\ D {\bf 7} (1973) 1279.

\bibitem{Estabrooks:1974vu}
  P.~Estabrooks and A.~D.~Martin,
  Nucl.\ Phys.\ B {\bf 79} (1974) 301.

\bibitem{Froggatt:1977hu}
  C.~D.~Froggatt and J.~L.~Petersen,
  Nucl.\ Phys.\ B {\bf 129} (1977) 89.

\bibitem{rho} 
  P.~Huovinen, P.~M.~Lo, M.~Marczenko, K.~Morita, K.~Redlich and C.~Sasaki,
  Phys.\ Lett.\ B {\bf 769}, 509 (2017).

\bibitem{exvol} 
  P.~M.~Lo, B.~Friman, M.~Marczenko, K.~Redlich and C.~Sasaki,
  Phys.\ Rev.\ C {\bf 96}, no. 1, 015207 (2017).





\bibitem{Kaplan:1998tg} 
  D.~B.~Kaplan, M.~J.~Savage and M.~B.~Wise,
  Phys.\ Lett.\ B {\bf 424}, 390 (1998).



\bibitem{Byckling:1971vca} 
E.~Byckling and K.~Kajantie,
\textit{Particle Kinematics} (John Wiley \& Sons Ltd, 1973).


\bibitem{Hormuzdiar:2000vq} 
  J.~Hormuzdiar, S.~D.~H.~Hsu and G.~Mahlon,
  Int.\ J.\ Mod.\ Phys.\ E {\bf 12}, 649 (2003).


\bibitem{Rischke:2001bt} 
  D.~H.~Rischke,
  Nucl.\ Phys.\ A {\bf 698}, 153 (2002).


\bibitem{Koch:2002uq} 
  V.~Koch,
  Nucl.\ Phys.\ A {\bf 715}, 108 (2003).


\bibitem{Becattini:2004td} 
  F.~Becattini,
  J.\ Phys.\ Conf.\ Ser.\  {\bf 5}, 175 (2005).


\bibitem{Pathria} 
R.~K.~Pathria, P.~D.~Beale,
\textit{Statistical Mechanics 3rd} (Academic Press, 2011).

\bibitem{Vasiliev:1998cq} 
A.~N.~Vasiliev,
\textit{Functional methods in quantum field theory and statistical physics} (Gordon and Breach, 1998).
    

\bibitem{Bedaque:2002xy} 
  P.~F.~Bedaque and G.~Rupak,
  Phys.\ Rev.\ B {\bf 67}, 174513 (2003).

\bibitem{Leyronas} 
  X.~Leyronas,
  Phys.\ Rev.\ A {\bf 84}, 053633 (2011).


\bibitem{Kaplan:2011br} 
  D.~B.~Kaplan and S.~Sun,
  Phys.\ Rev.\ Lett.\  {\bf 107}, 030601 (2011).
  doi:10.1103/PhysRevLett.107.030601

\bibitem{Liu}
X.~Liu
Phys.\ Rept.\  {\bf 524}, 2 (2013).

\bibitem{Lee:1952ig} 
  T.~D.~Lee and C.~N.~Yang,
  Phys.\ Rev.\  {\bf 87}, 410 (1952).

\bibitem{Ushcats1} 
  M.~V.~Ushcats
  Phys.\ Rev.\ Lett. {\bf 109}, 040601 (2012).

\bibitem{Ushcats2} 
  M.~V.~Ushcats
  Phys.\ Rev.\ E {\bf 91}, 052144 (2015).
  
\bibitem{Barnes:1991em} 
  T.~Barnes and E.~S.~Swanson,
  Phys.\ Rev.\ D {\bf 46}, 131 (1992).

\bibitem{Lacroix:2014gsa} 
  G.~Lacroix, C.~Semay and F.~Buisseret,
  Int.\ J.\ Mod.\ Phys.\ A {\bf 30}, no. 24, 1550145 (2015).

\bibitem{Briceno:2017tce} 
  R.~A.~Briceno, M.~T.~Hansen and S.~R.~Sharpe,
  Phys.\ Rev.\ D {\bf 95}, no. 7, 074510 (2017).

\bibitem{Danielewicz} 
  P.~Danielewicz and S.~Pratt,
  Phys.\ Rev.\ C {\bf 53}, 249 (1996).



\end{thebibliography}
\end{document}